\documentclass[conference]{IEEEtran}
\IEEEoverridecommandlockouts
\usepackage{amsmath,amssymb,amsfonts}
\usepackage{algorithmic}
\usepackage{graphicx}
\usepackage{textcomp}
\usepackage{xcolor}
\usepackage{booktabs}
\usepackage{subcaption}
\usepackage{makecell}
\usepackage{threeparttable}
\usepackage[backend=biber,style=ieee,citestyle=numeric]{biblatex}
\def\BibTeX{{\rm B\kern-.05em{\sc i\kern-.025em b}\kern-.08em
    T\kern-.1667em\lower.7ex\hbox{E}\kern-.125emX}}
\vspace{-0.8cm}
\begin{document}

\newcommand{\jian}[1]{\textcolor{red}{#1}} 

\title{SpNeRF: Memory Efficient Sparse Volumetric Neural Rendering Accelerator for Edge Devices\\

\thanks{* Corresponding author.}
\thanks{}
}
\author{\IEEEauthorblockN{Yipu Zhang$^1$, Jiawei Liang$^1$, Jian Peng$^1$, Jiang Xu$^2$, Wei Zhang$^{1,*}$}
\IEEEauthorblockA{$^1$\textit{Department of Electronic and Computer Engineering, The Hong Kong University of Science and Technology}\\
$^2$\textit{Microelectronics Thrust, The Hong Kong University of Science and Technology (GZ)}\\
\{yzhangqg, jliangbr, jpengai\}@connect.ust.hk, jiang.xu@hkust-gz.edu.cn, wei.zhang@ust.hk}
\vspace{-16pt}
}

\maketitle

\begin{abstract}
Neural rendering has gained prominence for its high-quality output, which is crucial for AR/VR applications. However, its large voxel grid data size and irregular access patterns challenge real-time processing on edge devices. While previous works have focused on improving data locality, they have not adequately addressed the issue of large voxel grid sizes, which necessitate frequent off-chip memory access and substantial on-chip memory.

This paper introduces SpNeRF, a software-hardware co-design solution tailored for sparse volumetric neural rendering. We first identify memory-bound rendering inefficiencies and analyze the inherent sparsity in the voxel grid data of neural rendering. To enhance efficiency, we propose novel preprocessing and online decoding steps, reducing the memory size for voxel grid. The preprocessing step employs hash mapping to support irregular data access while maintaining a minimal memory size. The online decoding step enables efficient on-chip sparse voxel grid processing, incorporating bitmap masking to mitigate PSNR loss caused by hash collisions. To further optimize performance, we design a dedicated hardware architecture supporting our sparse voxel grid processing technique. Experimental results demonstrate that SpNeRF achieves an average 21.07× reduction in memory size while maintaining comparable PSNR levels. When benchmarked against Jetson XNX, Jetson ONX, RT-NeRF.Edge and NeuRex.Edge, our design achieves speedups of 95.1×, 63.5×, 1.5× and 10.3×, and improves energy efficiency by 625.6×, 529.1×, 4×, and 4.4×, respectively. 
\end{abstract}

\vspace{-10pt}

\section{Introduction}


Neural Radiance Field (NeRF) \cite{mildenhall2020nerf} represents a novel and promising approach for 3D scene rendering, particularly attractive for Augmented and Virtual Reality (AR/VR) applications due to its high rendering quality and relatively high rendering speed on high-end GPUs. While state-of-the-art (SOTA) NeRF algorithms \cite{muller2022instant} achieve real-time performance on high-end  GPUs, they still struggle to meet the required processing speed on edge computing platforms, such as Jetson Xavier NX (XNX) \cite{ditty2018nvidia} and Jetson Orin NX (ONX) \cite{ditty2022nvidia}. The irregular memory access pattern introduced by multi-resolution hash encoding in \cite{muller2022instant} has been identified as a major efficiency bottleneck in both rendering and training. 


Several hardware-software co-design methods \cite{li2023instant,lee2023neurex,zhao2023instant,li2022rt} have been proposed to address rendering and training challenges in NeRF. Among these, Instant-3D \cite{li2023instant} and Instant-NeRF \cite{zhao2023instant} are ASIC accelerators focusing on improving data locality during NeRF training. Two latest ASIC neural rendering accelerators, NeuRex \cite{lee2023neurex} and RT-NeRF \cite{li2022rt}, exemplify different approaches to tackling rendering efficiency. NeuRex enhances data locality through restricted hashing, dividing large hash tables into subtables, but fails to address the large data size issue. RT-NeRF leverages sparsity in TensoRF \cite{chen2022tensorf} weight matrices using hybrid encoding. However, its performance is constrained by additional matrix-vector multiplications and limited exploration of voxel grid sparsity.

These approaches leave the large memory size problem unresolved, leading to \textbf{frequent off-chip memory access} and \textbf{large on-chip memory requirements}, significantly impeding existing works' overall performance. Recent algorithm research has explored model compression techniques, such as VQRF \cite{li2023compressing}, which aims to minimize memory size by identifying redundancies in voxel grid data and pruning less important points. While this approach shows promise in reducing model size, it introduces \textbf{new challenges} when deployed on edge computing platforms: 
\textbf{First}, the original VQRF consumes considerable memory during the rendering process. This is because VQRF adopts a method of restoring the full voxel grid from compressed data before processing, which leads to frequent off-chip memory access and increased latency, posing challenges in both power consumption and performance. To reduce the memory size, we propose a hash mapping based preprocessing, reducing the memory size significantly. \textbf{Second}, the irregularity introduced by ray sampling in VQRF complicates the process of locating voxel grid positions and fetching non-zero data from compressed encoded data. Existing encoding methods for Sparse Matrix Multiplication (SpMM) are not suitable for the irregular data access patterns in VQRF. To overcome this challenge, we integrate an online decoding step into our processing flow. \textbf{Third}, current edge computing platforms lack effective support for our proposed preprocessing and online decoding steps. To mitigate this issue, we design a dedicated hardware architecture optimized for these operations, ensuring peak performance. 

\begin{figure*}
    \centering
    \includegraphics[width=\linewidth]{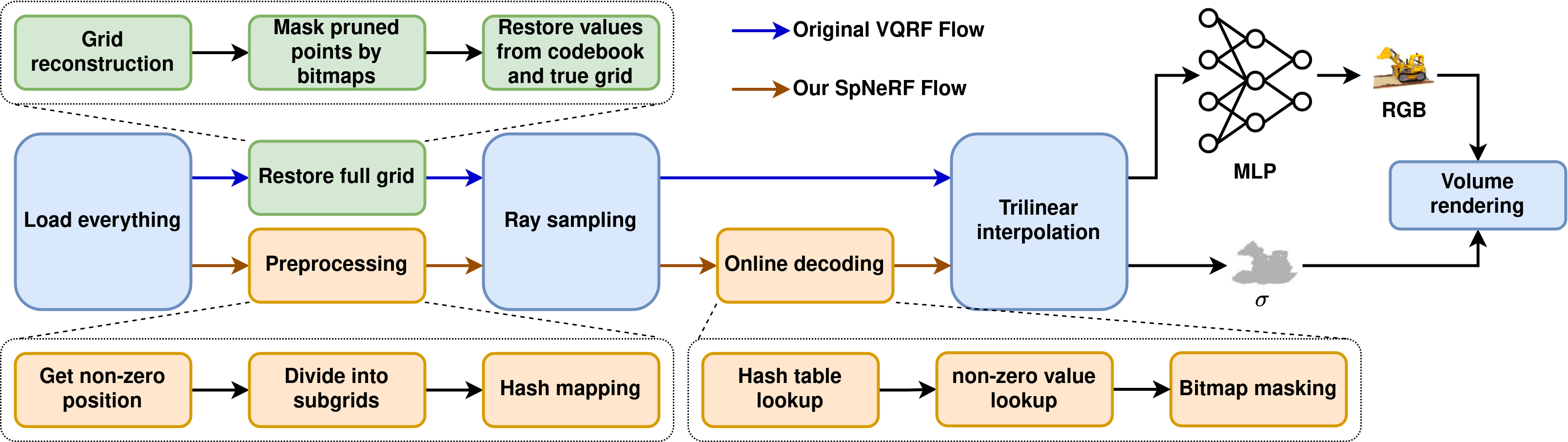}
    \caption{Orginal VQRF flow and our SpNeRF flow}
    \label{fig:flow}
    \vspace{-12pt}
\end{figure*}

To summarize, this work makes the following contributions:
\begin{itemize}
\item We propose SpNeRF, a software-hardware co-design framework leveraging sparsity in neural rendering to overcome the memory-bound bottleneck faced by previous works.
\item We introduce a preprocessing step utilizing hash mapping to enhance memory mapping efficiency for sparse voxel grids, minimizing the memory size for the voxel grid and simplifying non-zero data location in compressed structures during irregular memory access.
\item We develop an online decoding step incorporating hash table lookup, non-zero value lookup, and bitmap masking, mitigating accuracy loss caused by hash collisions.
\item We design a dedicated hardware architecture integrating a Sparse Grid Processing Unit (SGPU) and MLP Unit to efficiently support our SpNeRF algorithm, boosting sparse volumetric neural rendering efficiency.
\end{itemize}

    
\section{Background and Motivation}

\subsection{Vanilla NeRF and VQRF}


 Vanilla NeRF \cite{mildenhall2020nerf} has revolutionized novel view synthesis by integrating computer graphics techniques with deep learning. As a pioneering work, NeRF introduced the use of deep neural networks to represent and render 3D scenes from a sparse set of input views. However, the original NeRF implementation relies on computationally expensive multi-layer perceptrons (MLPs) with a large number of parameters. Consequently, the rendering process can be time-consuming, often requiring hours or even days to generate a single novel view. Therefore, the volumetric method is proposed to boost training and rendering efficiency \cite{liu2020neural}. It estimated the color features and density of sampling points by interpolating the data stored in the voxel grid.

 VQRF \cite{li2023compressing}, as shown in Fig. \ref{fig:flow}, identifies the redundancy in voxel grid data and proposes voxel pruning and vector quantization to minimize the memory size. Also, VQRF has a relatively smaller MLP (only 3 layers with channel sizes of 128, 128, 3). However, it cannot fully take advantage of the sparsity brought by the voxel pruning since it requires restoring the full voxel grid using the pruned voxel grid points. Therefore, it leads to low efficiency when rendering on edge computing platforms, such as XNX and ONX.

\subsection{Encoding Method}
Various encoding methods have been proposed for SpMM to enhance memory and computation efficiency \cite{dave2021hardware}. However, current approaches have limitations. The COO format requires storing all coordinates, which brings an extra 630 KB of memory usage for each scene on average in our experiments. CSR and CSC formats necessitate row-wise or column-wise data storage, respectively. While COO offers simple implementation but high memory overhead, CSR provides efficient row-wise access at the cost of poor column-wise performance, and CSC excels in column-wise operations but struggles with row-wise access. These encoding methods often result in excessive memory consumption and numerous lookups during irregular data access. Consequently, these factors lead to frequent off-chip memory accesses and the large on-chip memory demand, both of which significantly impact performance. A detailed analysis of these performance implications will be presented in Section \ref{profile}.

\subsection{Profiling VQRF on GPUs}\label{profile}

\begin{figure}
    \centering
    
    \begin{minipage}[b]{0.24\textwidth} 
        \centering
        \includegraphics[width=\textwidth]{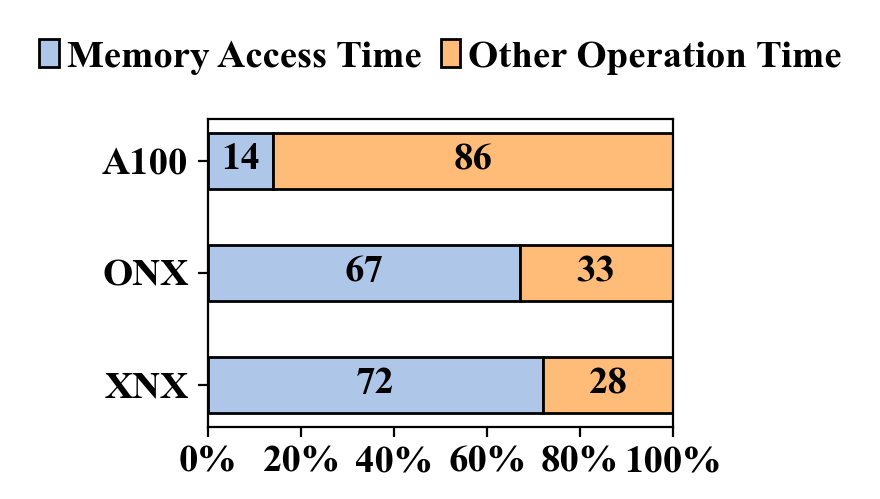}
        \subcaption{Time distribution}
        \label{fig:profile}
    \end{minipage}
    \begin{minipage}[b]{0.24\textwidth} 
        \centering
        \includegraphics[width=\textwidth]{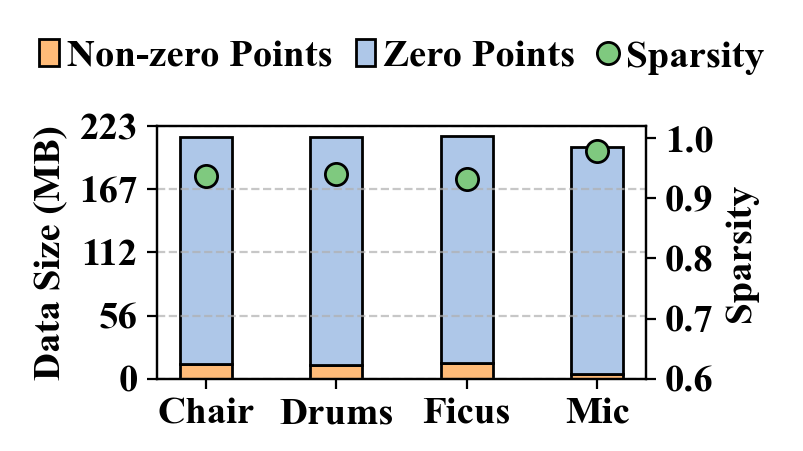}
        \subcaption{Voxel grid data sparsity}
        \label{fig:sparsity}
    \end{minipage}
    \caption{Profiling result for runtime on GPUs and sparsity on datasets }
    \label{fig:motivation}
\end{figure}

To identify VQRF bottlenecks, we profile the neural rendering process across various computing platforms and characterize sparse data in different datasets. We utilize one high-end computing platform, NVIDIA A100 (A100), and two edge computing platforms, ONX, and XNX, whose specifications are summarized in Table \ref{tab:Spec}. Evaluating VQRF on Synthetic-NeRF \cite{mildenhall2020nerf} datasets for each GPU, we present the runtime breakdown in Fig. \ref{fig:motivation}(\subref{fig:profile}). Profiling results reveal that edge computing platforms spend the most time accessing memory, while the A100 allocates minimal time to this task. We also examine voxel grid data redundancy by separating non-zero and zero components and calculate sparsity, as illustrated in Fig. \ref{fig:motivation}(\subref{fig:sparsity}).
\begin{table}[htbp]
    \centering
    \caption{A summary of profiling computing platforms}
    \begin{tabular}{c|c c c}
    \hline
     Spec. & A100 \cite{choquette2020nvidia} & ONX \cite{ditty2022nvidia} & XNX \cite{ditty2018nvidia} \\
     \hline
     Tech. & 7 nm & 8 nm & 16 nm\\
     \hline
     Power & 400 W & 25 W & 20 W \\
     \hline
     DRAM &\makecell{5120-bit 40 GB\\ HBM2\\ 1555 GB/s} & \makecell{128-bit 16 GB\\ LPDDR5\\ 102.4 GB/s} & \makecell{128-bit 16 GB\\ LPDDR4\\ 59.7 GB/s} \\
     \hline
     GPU L2 cache & 40 MB & 4 MB & 512 KB\\
     \hline
     FP32 & 19.5 TFLOPS & 1.9 TFLOPS & 885 GFLOPS \\
     \hline
     FP16 & 78 TFLOPS & 3.8 TFLOPS & 1.69 TFLOPS \\
     \hline
\end{tabular}
    \label{tab:Spec}
\end{table}

Our profiling results yield two key observations: \textbf{First}, the rendering process on edge devices is predominantly memory bandwidth-bound. The proportion of time spent on memory access in edge computing platforms is $4.79\times \sim 5.14\times$ higher than in high-end computing platforms, due to the smaller L2 cache size and relatively low DRAM bandwidth of edge devices. \textbf{Second}, there is substantial redundancy in voxel grid data. 
As illustrated in Fig. \ref{fig:motivation}(\subref{fig:sparsity}), non-zero points occupy only $2.01\% \sim 6.48\%$ of total voxel grid data, indicating significant potential for leveraging sparsity to enhance efficiency. These findings underscore the need for an on-chip processing flow to mitigate expensive off-chip memory access and alleviate the memory bandwidth bottleneck.


\section{Algorithm Design}
In this section, we introduce the SpNeRF algorithm design, which enables efficient on-chip processing by fully exploiting sparsity to address the memory-bound problem. Our method replaces the conventional restore step with a preprocessing step and incorporates an online decoding step, which is shown in Fig. \ref{fig:flow}, significantly reducing memory bandwidth requirements and enabling efficient on-chip computation. 

\begin{figure}
    \centering
    \includegraphics[width=\linewidth]{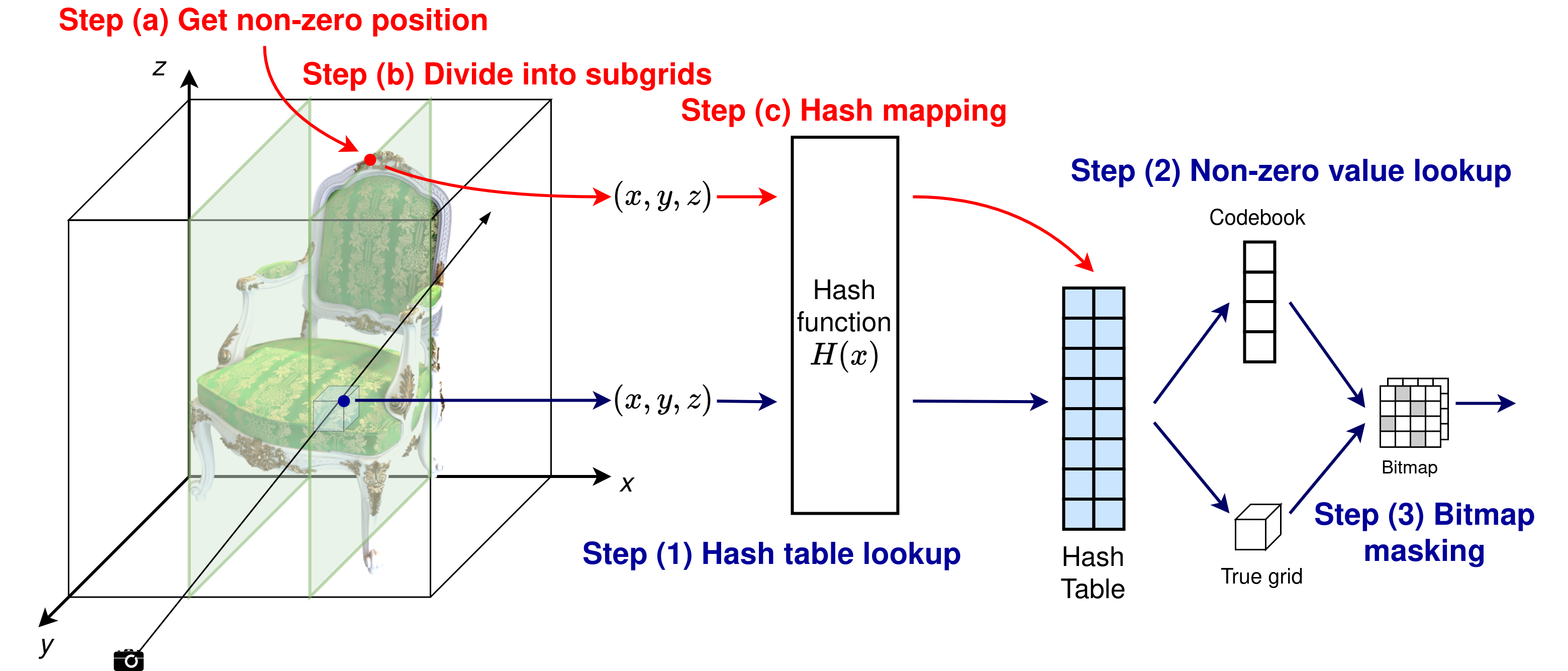}
    \caption{An illustration of preprocessing and online decoding flow}
    \label{fig:preprocess}
\end{figure}

\subsection{Hash Mapping Based Preprocessing}\label{AA}

The preprocessing step encompasses three stages, as shown in the red line in Fig. \ref{fig:preprocess}. Initially, we identify the non-zero points in the sparse voxel grid and extract their position coordinates $(x,y,z)$ into a vector $\boldsymbol{p} = [x, y, z]^T$. We then aggregate all these position vectors into a set $P_{nz} = \{\boldsymbol{p}_i | i = 1, 2, ..., N\}$, where $N$ is the total number of non-zero points. This step preserves the spatial information and saves it for further voxel grid partition and hash mapping.
\begin{figure}[htbp]
    \centering
    \includegraphics[width=\linewidth]{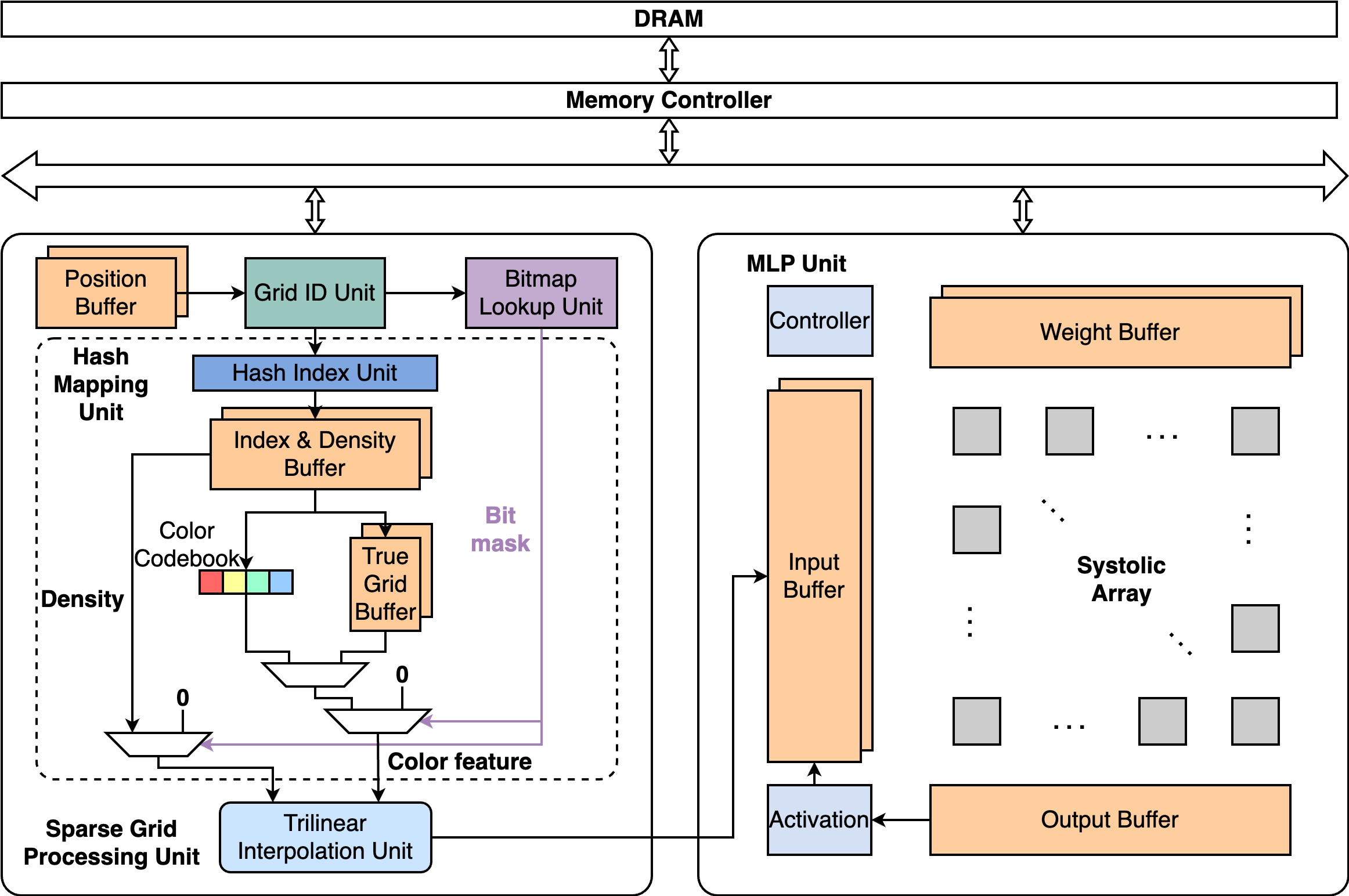}
    \caption{SpNeRF accelerator architecture}
    \label{fig:arch}
\end{figure}

Subsequently, we partition the identified non-zero points into $K$ subgrids based on their $x$ coordinate values. This partitioning is defined as $S_k = \{\boldsymbol{p}_i | \lfloor x_i / w \rfloor = k, \boldsymbol{p}_i \in P_{nz}\}$, where $k \in {0, 1, ..., K-1}$, $w$ is the width of each subgrid, and $x_i$ is the $x$-coordinate of $\boldsymbol{p}_i$.

Consequently, to efficiently support the irregular data access patterns inherent in neural rendering, we map each subgrid $S_k$ into a separate hash table $H_k$ using the following function from \cite{muller2022instant}:
\begin{equation}
    h(\boldsymbol{p_i}) = (x_i \pi_1 \oplus y_i \pi_2 \oplus z_i \pi_3)\mod{T}\label{eq:hash}
\end{equation}
where $T$ is the number of entries per hash table level, $\pi_1 = 1$, $\pi_2 = 2654435761$, and $\pi_3 = 805459861$. Note that each hash table entry stores the index for non-zero value lookup in online decoding.
This final mapping process enables rapid lookup during the rendering process and eliminates the need for storing coordinates, thus solving the memory-bound problem and optimizing overall performance. 

\begin{figure*}[htbp]
    \centering
    \includegraphics[width=.9\linewidth]{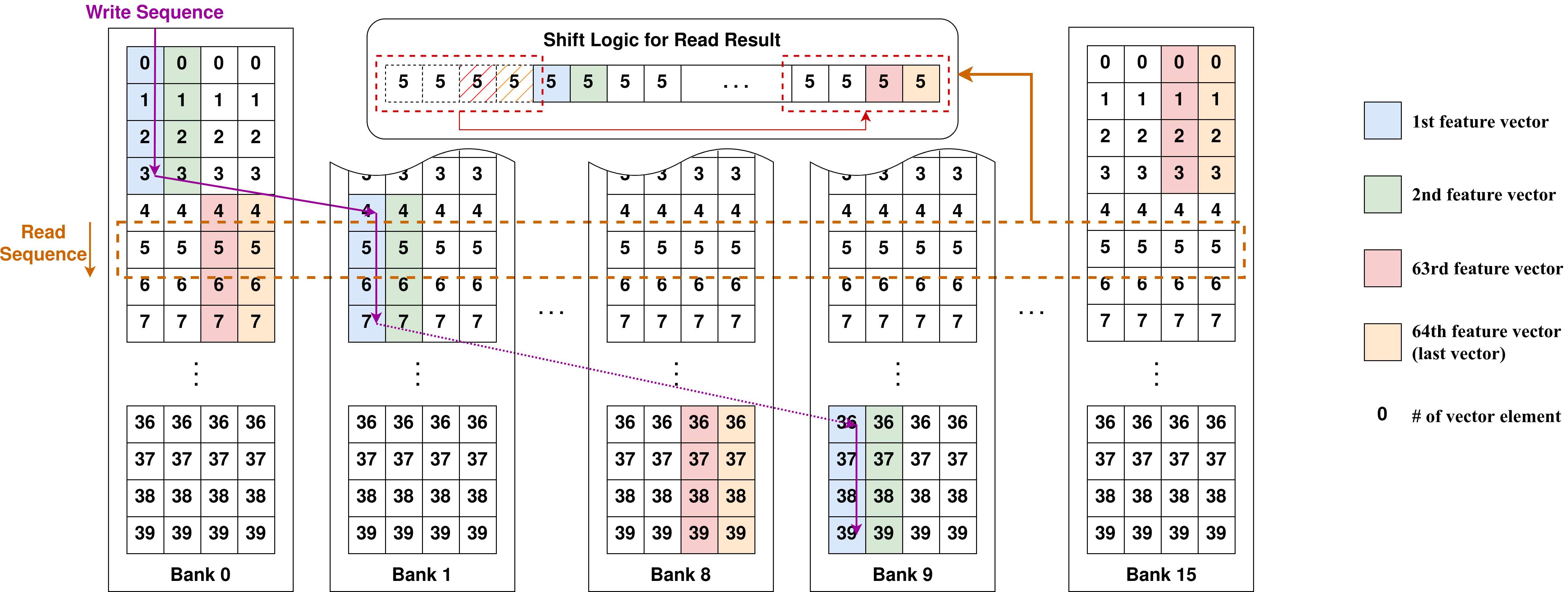}
    \caption{Our proposed block-circulant storage format for input buffer}
    \label{fig:MLP_buffer}
    \vspace{-12pt}
\end{figure*}

\subsection{Online Sparse Voxel Grid Decoding}\label{hash mapping}
We incorporate an online decoding step between ray sampling and interpolation to support efficient data retrieval, as illustrated by the blue line in Fig. \ref{fig:preprocess}. This online decoding process unfolds as follows: Initially, for each sample point, we retrieve its position and calculate the corresponding hash index using \eqref{eq:hash}. Subsequently, we utilize this hash index to fetch the lookup index for non-zero values from the hash table. Following this, we employ the retrieved 18-bit index to locate non-zero values in both the codebook and true voxel grid. To streamline this process, we implement a unified 18-bit addressing scheme for both the codebook and true voxel grid. Finally, we implement a bitmap masking technique to mitigate errors primarily caused by hash collisions. This approach utilizes a bitmap that stores a single bit for each voxel grid point, indicating whether the point is zero (0) or non-zero (1). By representing all voxel grid points' status in just 1 bit each, the bitmap provides a memory-efficient method to track non-zero values. During the decoding process, we consult this bitmap to effectively mask all erroneous values resulting from hash collisions, setting them to zero. Our observations indicate that hash collisions are the dominant source of errors in this process. Therefore, the bitmap masking step is crucial for maintaining accuracy in our decoding procedure. The detailed result for bitmap masking will be analyzed in Section \ref{sec:alg_eva}.

\section{Hardware Architecture Design}
\subsection{Overview}
In this section, we present the SpNeRF architecture as illustrated in Fig. \ref{fig:arch}, designed to support our algorithm. The SpNeRF accelerator comprises two primary modules: the Sparse Grid Processing Unit (SGPU) and the MLP Unit. The SGPU is specifically tailored to execute the series of lookup operations required during online sparse voxel grid decoding. Complementing the SGPU, the MLP Unit is implemented as an output-stationary systolic array, a design commonly employed in conventional DNN accelerators. Our design method aligns with \cite{lee2023neurex}, extending a standard DNN accelerator with a specialized unit (in our case, the SGPU) to efficiently support neural rendering tasks. The overall dataflow of our system is initiated at the position buffer. Upon retrieval of a position, it is forwarded to the Hash Mapping Unit, which performs a lookup operation to locate the corresponding voxel grid data. The Trilinear Interpolation Unit subsequently processes this voxel grid data to compute the interpolation result. The resulting output is then concatenated with the view direction vector and stored in the input buffer of the MLP unit. The on-chip computing is in FP16 while the true voxel grid data is saved in INT8 format on off-chip memory to save the memory usage and reduce the communication overhead. To ensure high throughput, the entire design is fully pipelined. Furthermore, all buffers in the system are double-buffered, enabling simultaneous data fetching and processing. 

\begin{figure*}[htbp]
    \centering
    
    \begin{minipage}[b]{0.48\textwidth} 
        \centering
        \includegraphics[width=\textwidth]{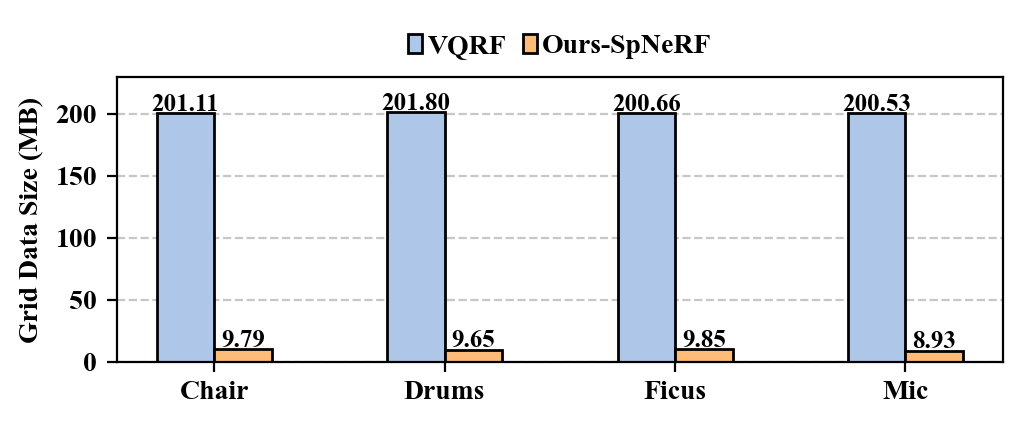}
        \subcaption{Memory size reduction}
        \label{fig:mem_size}
    \end{minipage}
    \begin{minipage}[b]{0.48\textwidth} 
        \centering
        \includegraphics[width=\textwidth]{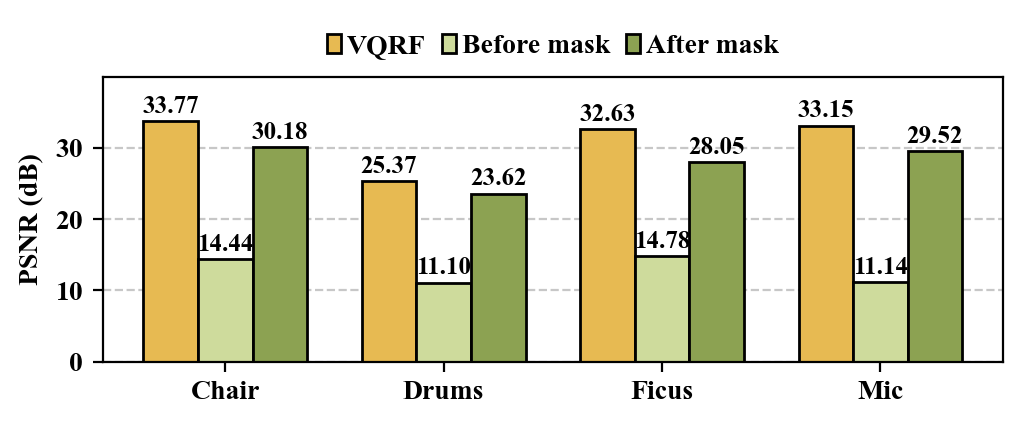}
        \subcaption{PSNR result}
        \label{fig:PSNR}
    \end{minipage}
    \caption{Algorithm evaluation result for our proposed SpNeRF }
    \label{fig:alg_eva}
    \vspace{-12pt}
\end{figure*}
\subsection{Sparse Grid Processing Unit}

The Sparse Grid Processing Unit (SGPU) is designed to support our proposed online sparse voxel grid decoding flow. It comprises four key components:

\textbf{Grid ID Unit (GID).} The GID computes ceiling and round results for each point position to locate the corresponding voxel grid vertex. It also calculates the weight by FP16 multipliers and subtractors for trilinear interpolation using the following equation introduced by \cite{lee2023neurex}:
\begin{equation}
w = (1 - |x_p - x_g| )\cdot(1-|y_p-y_g|)\cdot(1-|z_p-z_g|)\label{eq:interp}
\end{equation}
where $(x_p,y_p,z_p)$ represents the sample point position and $(x_g,y_g,z_g)$ denotes the voxel grid vertex position.

\textbf{Bitmap Lookup Unit (BLU).} The BLU stores the bit mask for the current subgrid. All bit masks are stored sequentially in a contiguous memory space, enabling efficient location of bits using voxel grid vertex positions as memory addresses and selection signals. The BLU lookup results are used to mask incorrect non-zero values caused by hash collisions in the Hash Mapping Unit.

\textbf{Hash Mapping Unit (HMU).} The HMU is the core module of the SGPU, facilitating the hash table lookup crucial for our online sparse voxel grid decoding. It processes the voxel grid vertex results from the GID and computes the hash index using Equation \eqref{eq:hash}. Once the hash index is determined, the color feature index and density are fetched from the hash table stored in the Index and Density Buffer. Our unified 18-bit addressing scheme differentiates color codebook and true voxel grid buffer lookups by comparing the index value. For a color codebook size of $4096\times 12$, lookup requests with color feature index values below 4096 are served by the color codebook, while others are directed to the true voxel grid buffer. The final output is filtered by the bit mask from BLU.

\textbf{Trilinear Interpolation Unit (TIU).} The TIU plays a crucial role in processing and interpolating color features. Initially, it converts the original color features from the true voxel grid buffer, stored in INT8 format, to FP16 format by multiplying the lookup results with the scale factor. Meanwhile, the density data and the color feature vector from the codebook will be directly sent to later computations. Following this conversion, the TIU multiplies each transformed color feature with its corresponding voxel grid vertex weight, as computed by the voxel grid ID Unit. Finally, it accumulates the weighted color features from all eight surrounding voxel grid vertices to produce the final interpolation result. This process can be represented by the equation $C_{interp} = \sum_{i=1}^8 w_i \cdot (s \cdot C_i)$, where $C_{interp}$ is the interpolation result, $w_i$ are the weights for each voxel grid vertex, $s$ is the scale factor for de-quantization, and $C_i$ color feature vector in corresponding voxel grid vertex.

\subsection{MLP Unit}

The MLP Unit consists of an output-stationary systolic array, activation unit, controller unit, and data buffers. It computes a 3-layer MLP with channel sizes of 128, 128, and 3, respectively. To enhance efficiency, we implement batch processing with a batch size of 64. Addressing the challenge of inconsistent vector sizes between the previous unit's output and the systolic array's input dimension, we propose a block-circulant storage format. This optimized memory pattern reduces both memory overhead and read time.

As illustrated in Fig. \ref{fig:MLP_buffer}, the block-circulant storage format interleaves the input $39\times 1$ vector across banks 0 to 9. Each input vector is segmented into 10 blocks, with each block containing four consecutive elements. Elements within the same block are stored successively, while adjacent blocks are stored in neighboring banks with a block offset of 4. The write sequence depicted in Fig. \ref{fig:MLP_buffer} demonstrates the process of writing the first vector into the input buffer. To ensure divisibility by 4, we pad the last element with 0. The reading sequence follows the vector element number in ascending order, with each read result undergoing a block shift to maintain the correct sequence of the first vector. As shown in Fig. \ref{fig:MLP_buffer}, the current reading vector passes through shift logic, which repositions the first block from bank 0 to ensure that elements of the first vector are correctly aligned with the first row of the systolic array.

\section{Evaluation}

\subsection{Evaluation Setup}


\textbf{Implementation.} For \underline{Algorithm Evaluation}, we implement our proposed SpNeRF algorithm based on VQRF \cite{li2023compressing} in PyTorch \cite{paszke2019pytorch}. For \underline{Hardware Evaluation}, we implement our accelerator in Verilog and synthesize our design using Synopsys Design Compiler based on TSMC 28nm CMOS Technology to obtain the power and area metrics. The operating clock frequency for our design is 1 GHz. On-chip SRAMs are generated by the provided memory compiler with the same technology. To evaluate the overall performance of our SpNeRF architecture, we develop a cycle-level simulator verified against our RTL design. The DRAM timing and power characteristics are obtained using Ramulator \cite{kim2015ramulator} with the configuration of LPDDR4-3200, which provides a bandwidth of 59.7 GB/s.

\begin{figure}
    \centering
    
    \begin{minipage}[b]{0.24\textwidth} 
        \centering
        \includegraphics[width=\textwidth]{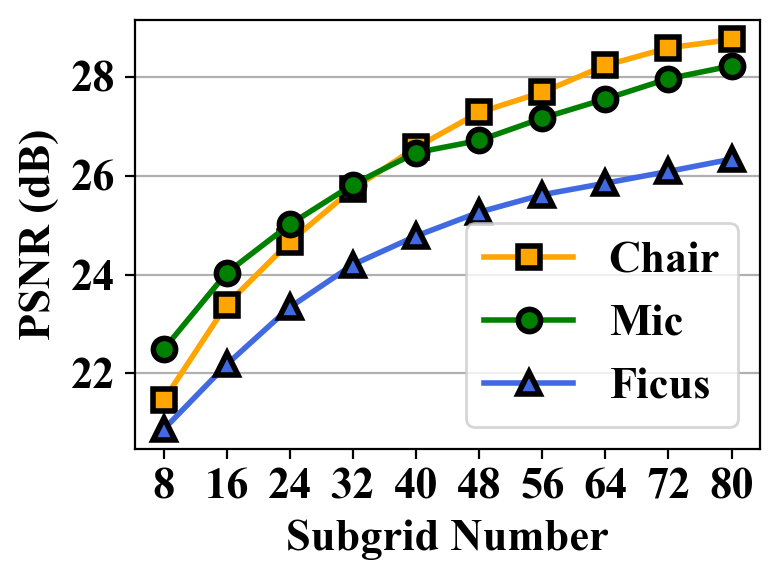}
        \subcaption{Hash table size = 16 k}
        \label{fig:subgrid}
    \end{minipage}
    \begin{minipage}[b]{0.24\textwidth} 
        \centering
        \includegraphics[width=\textwidth]{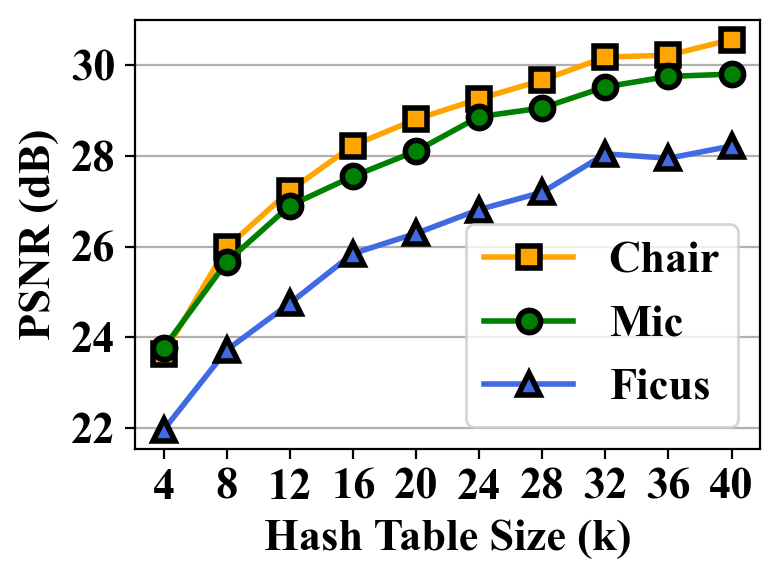}
        \subcaption{Subgrid number = 64}
        \label{fig:table_size}
    \end{minipage}
    \caption{(a) PSNR vs. different subgrid number and (b) PSNR vs. hash table size}
    \label{fig:relation}
\end{figure}
\textbf{Datasets \& Baseline.} To evaluate the performance of the proposed SpNeRF, we conduct experiments on the Synthetic-NeRF \cite{mildenhall2020nerf} dataset. For the hardware evaluation baseline, we select the edge computing platforms and edge accelerators as our baselines. For edge computing platforms, we compare our design with Jetson Xavier NX 16 GB and Jetson Orin NX 16 GB (commonly used edge GPUs). For edge accelerators, we compare our design with RT-NeRF.edge and NeuRex.edge (both are dedicated ASIC accelerators for neural rendering).
\subsection{Algorithm Evaluation}\label{sec:alg_eva}

\textbf{Subgrid Number \& Hash table size.}
Fig. \ref{fig:relation} illustrates the relationship among PSNR, subgrid number, and hash table size. The results reveal that PSNR increases rapidly initially but slow down beyond certain values for both subgrid number and hash table size. Based on this analysis, our design adopts a subgrid number of 64 and a hash table size of 32 k, as larger values yield only marginal improvements in PSNR.
\begin{figure*}
    \centering
    
    \begin{minipage}[b]{0.48\textwidth} 
        \centering
        \includegraphics[width=\textwidth]{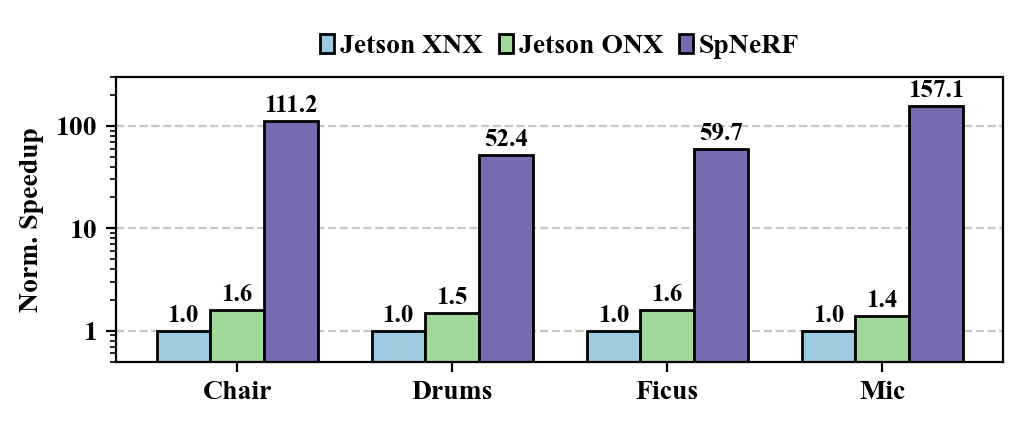}
        \subcaption{Normalized speedup compared with XNX and ONX}
        \label{fig:speed}
    \end{minipage}
    \begin{minipage}[b]{0.48\textwidth} 
        \centering
        \includegraphics[width=\textwidth]{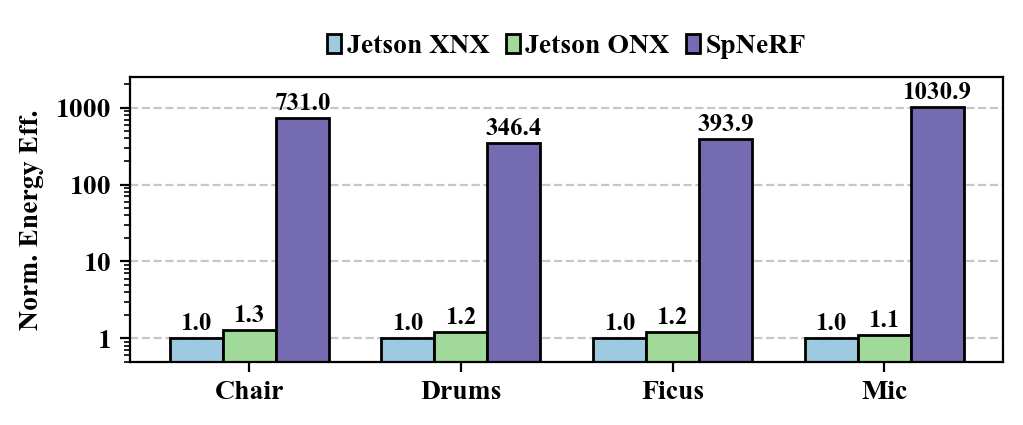}
        \subcaption{Normalized energy efficiency compared with XNX and ONX}
        \label{fig:eff}
    \end{minipage}
    \caption{Normalized speedup and energy efficiency compared with edge computing platforms}
    \label{fig:result}
    \vspace{-12pt}
\end{figure*}

\textbf{Memory Size Reduction.}
Fig. \ref{fig:alg_eva}(\subref{fig:mem_size}) compares the voxel grid data size of our proposed SpNeRF with the original VQRF. SpNeRF achieves an average memory size reduction of $21.07\times$ compared to VQRF. This significant reduction in memory footprint is primarily attributed to the online sparse voxel grid decoding technique, which eliminates the necessity of restoring the full voxel grid data.

\textbf{PSNR Performance.}
Fig. \ref{fig:alg_eva}(\subref{fig:PSNR}) illustrates the PSNR results for VQRF, SpNeRF before bitmap masking, and SpNeRF after bitmap masking. Higher PSNR values indicate better image quality. The results demonstrate that our SpNeRF, with bitmap masking applied, maintains PSNR levels comparable to VQRF while simultaneously achieving substantial memory reduction. This indicates that the proposed method effectively preserves image quality despite the significant decrease in memory usage.

\subsection{Hardware Evaluation}
\begin{figure}
    \centering
    
    \begin{minipage}[b]{0.25\textwidth} 
        \centering
        \includegraphics[width=\textwidth]{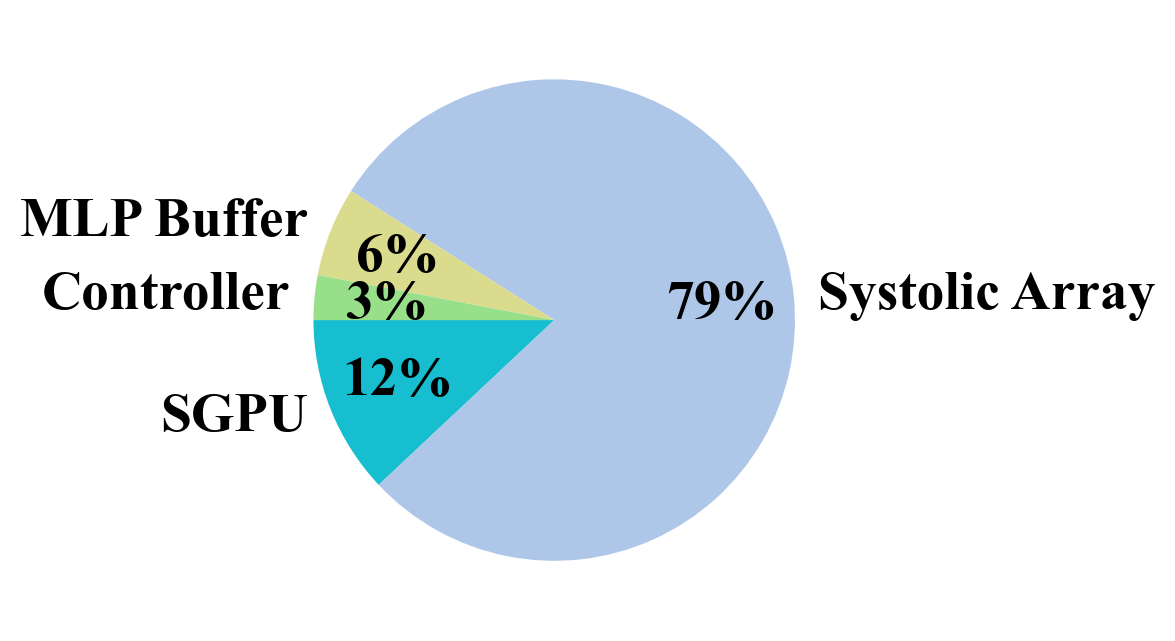}
        \subcaption{Area breakdown}
        \label{fig:area}
    \end{minipage}
    \begin{minipage}[b]{0.23\textwidth} 
        \centering
        \includegraphics[width=\textwidth]{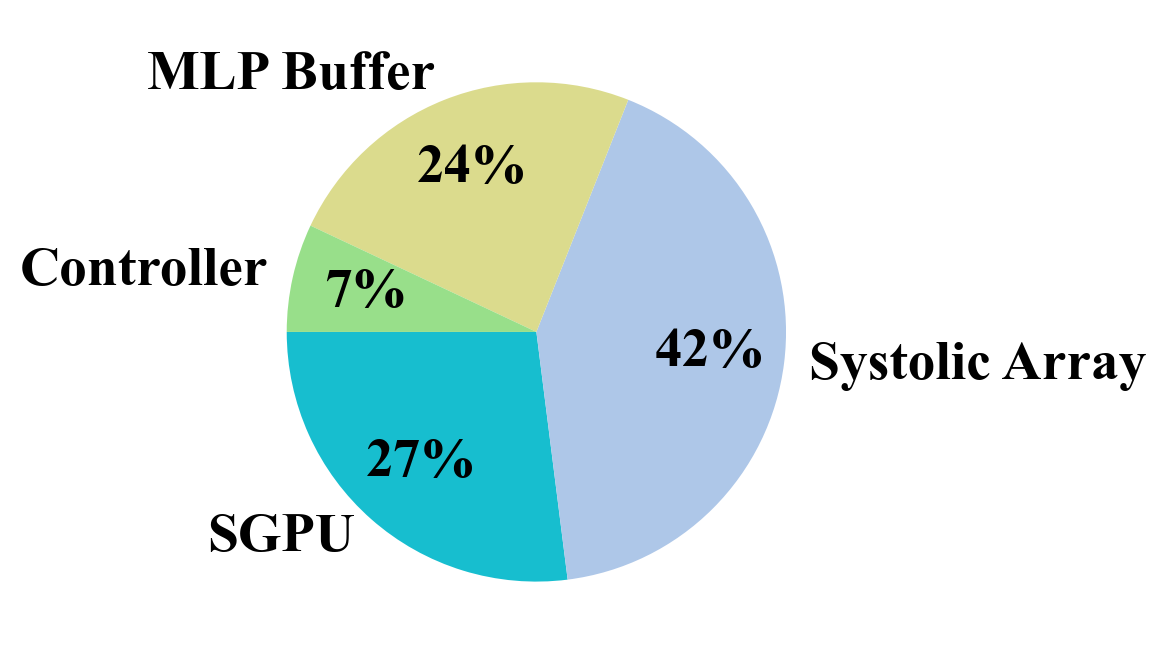}
        \subcaption{Power breakdown}
        \label{fig:power}
    \end{minipage}
    \caption{Area and Power of SpNeRF }
    \label{fig:hw_eva}
\end{figure}

\textbf{Performance.}
Fig. \ref{fig:result} shows the normalized speedup and energy efficiency compared with edge devices and Table \ref{tab:Compare} gives the comparison between our design and previous works. The results indicate that SpNeRF achieves great advances in both speedup and energy efficiency. In comparison, SpNeRF achieves $52.4\times \sim 157.1\times$ speedup compared with XNX and $34.9\times \sim 112.2\times$ speedup compared with ONX. This improvement is attributed to hash mapping adopted in preprocessing and online decoding, eliminating the need for frequent off-chip memory access.
\textbf{Area.}
Fig. \ref{fig:hw_eva}(\subref{fig:area}) illustrates the area breakdown of our SpNeRF accelerator, while Table \ref{tab:Compare} presents the total area and area efficiency of our design. In the area breakdown, the MLP buffer accounts for 58 KB SRAM (comprising input buffer, output buffer, and weight buffer), and the SGPU contains 571 KB SRAM. Notably, in our design, on-chip SRAM occupies only a small fraction of the overall area. This contrasts with other designs, where on-chip SRAM typically dominates the chip area. The results demonstrate that the memory reduction achieved through online sparse voxel grid decoding effectively minimizes on-chip SRAM size, leading to superior area efficiency of $2.67\times$ to $3.04\times$ compared to previous works.
\begin{table}[htbp]
    \centering
    \caption{Summary of comparisons between related work and our SpNeRF}
    \begin{threeparttable}
    \begin{tabular}{c|c c c}
    \Xhline{1px}
     Accelerator & RT-NeRF \cite{li2022rt} & NeuRex \cite{lee2023neurex} & SpNeRF (Ours) \\
     \hline
     SRAM ($MB$)   & 3.5  & 0.86  & 0.61  \\
     \hline
     Area ($mm^2$)   & 18.85  & 1.31  & 7.7  \\
     \hline
     Tech. ($nm$) & 28  & 28  & 28 \\
     \hline
     Power ($W$) & 8  & 1.31  & 3  \\
     \hline
     DRAM &\makecell{LPDDR4-1600\\ 17 GB/s} & \makecell{LPDDR4-3200 \\ 59.7 GB/s} & \makecell{LPDDR4-3200 \\ 59.7 GB/s} \\
     \hline
     FPS & 45 & 6.57\tnote{*} & \textbf{67.56} \\
     \hline
     \makecell{Energy Eff. \\ ($FPS/W$)} & 5.63  & 5.15  &\textbf{22.52 } \\
     \hline
     \makecell{Area Eff. \\ ($FPS/mm^2$)} & 2.38  & 2.09  & \textbf{6.36 } \\
     \Xhline{1px}
\end{tabular}
\begin{tablenotes}
\footnotesize
\item[*] NeuRex only provides normalized speedup. Here we infer from Jetson XNX rendering speed.
\end{tablenotes}
\end{threeparttable}
    \label{tab:Compare}
\end{table}

\textbf{Power.}
Fig. \ref{fig:hw_eva}(\subref{fig:power}) depicts the power breakdown of our SpNeRF accelerator and Table \ref{tab:Compare} provides the total power and power efficiency of our design. The power breakdown reveals that the systolic array accounts for the dominant portion of overall power consumption in our design, contrasting with previous studies where on-chip SRAM was the primary power consumer. Our approach achieves a $4\times$ to $4.37\times$ improvement over previous works. When comparing with XNX and ONX, as shown in Fig.\ref{fig:result}(\subref{fig:eff}), SpNeRF achieves $346.4\times \sim 1030.9\times$ and $288.7\times \sim 937.2\times$ energy efficiency improvement, respectively. This advantage comes from two key factors: First, our design minimizes on-chip SRAM size, which has been observed to be a dominant contributor to power consumption in previous works. Second, our design fully leverages voxel grid sparsity, significantly reducing off-chip memory access, another major source of power consumption.
\section{Conclusion}


This paper presents SpNeRF, a novel software-hardware co-design approach for facilitating memory efficiency in sparse volumetric neural rendering. We propose hash mapping-based preprocessing and online sparse voxel grid decoding techniques to fully leverage the sparsity in voxel grid data. To support our algorithmic innovations, we introduce a dedicated hardware architecture designed for efficient neural rendering. Experimental results demonstrate that our design significantly outperforms edge computing platforms and previous works in terms of speedup, energy efficiency, and area efficiency. These improvements make SpNeRF a promising solution for advancing the field of neural rendering, particularly in resource-constrained environments.

\section*{Acknowledgment}

This work was partially supported by AI Chip Center for Emerging Smart Systems (ACCESS), Hong Kong SAR and Collaborative Research Fund (UGC CRF) C5032-23G.

\newpage
\AtNextBibliography{\small}
\printbibliography{}

@inproceedings{mildenhall2020nerf,
  title={NeRF: Representing Scenes as Neural Radiance Fields for View Synthesis},
  author={Mildenhall, Ben and Srinivasan, Pratul P and Tancik, Matthew and Barron, Jonathan T and Ramamoorthi, Ravi and Ng, Ren},
  booktitle={European Conference on Computer Vision},
  pages={405--421},
  year={2020},
  organization={Springer}
}

@article{muller2022instant,
  title={Instant neural graphics primitives with a multiresolution hash encoding},
  author={M{\"u}ller, Thomas and Evans, Alex and Schied, Christoph and Keller, Alexander},
  journal={ACM transactions on graphics (TOG)},
  volume={41},
  number={4},
  pages={1--15},
  year={2022},
  publisher={ACM New York, NY, USA}
}

@inproceedings{ditty2018nvidia,
  title={Nvidia’s xavier soc},
  author={Ditty, Michael and Karandikar, Ashish and Reed, David},
  booktitle={Hot chips: a symposium on high performance chips},
  year={2018}
}

@inproceedings{ditty2022nvidia,
  title={Nvidia orin system-on-chip},
  author={Ditty, Michael},
  booktitle={2022 IEEE Hot Chips 34 Symposium (HCS)},
  pages={1--17},
  year={2022},
  organization={IEEE Computer Society}
}

@inproceedings{li2023instant,
  title={Instant-3d: Instant neural radiance field training towards on-device ar/vr 3d reconstruction},
  author={Li, Sixu and Li, Chaojian and Zhu, Wenbo and Yu, Boyang and Zhao, Yang and Wan, Cheng and You, Haoran and Shi, Huihong and Lin, Yingyan},
  booktitle={Proceedings of the 50th Annual International Symposium on Computer Architecture},
  pages={1--13},
  year={2023}
}

@inproceedings{lee2023neurex,
  title={Neurex: A case for neural rendering acceleration},
  author={Lee, Junseo and Choi, Kwanseok and Lee, Jungi and Lee, Seokwon and Whangbo, Joonho and Sim, Jaewoong},
  booktitle={Proceedings of the 50th Annual International Symposium on Computer Architecture},
  pages={1--13},
  year={2023}
}

@inproceedings{zhao2023instant,
  title={Instant-NeRF: Instant On-Device Neural Radiance Field Training via Algorithm-Accelerator Co-Designed Near-Memory Processing},
  author={Zhao, Yang Katie and Wu, Shang and Zhang, Jingqun and Li, Sixu and Li, Chaojian and Lin, Yingyan Celine},
  booktitle={2023 60th ACM/IEEE Design Automation Conference (DAC)},
  pages={1--6},
  year={2023},
  organization={IEEE}
}

@inproceedings{li2023compressing,
  title={Compressing volumetric radiance fields to 1 mb},
  author={Li, Lingzhi and Shen, Zhen and Wang, Zhongshu and Shen, Li and Bo, Liefeng},
  booktitle={Proceedings of the IEEE/CVF Conference on Computer Vision and Pattern Recognition},
  pages={4222--4231},
  year={2023}
}

@inproceedings{li2022rt,
  title={RT-NeRF: Real-time on-device neural radiance fields towards immersive AR/VR rendering},
  author={Li, Chaojian and Li, Sixu and Zhao, Yang and Zhu, Wenbo and Lin, Yingyan},
  booktitle={Proceedings of the 41st IEEE/ACM International Conference on Computer-Aided Design},
  pages={1--9},
  year={2022}
}

@inproceedings{choquette2020nvidia,
  title={Nvidia a100 gpu: Performance \& innovation for gpu computing},
  author={Choquette, Jack and Gandhi, Wish},
  booktitle={2020 IEEE Hot Chips 32 Symposium (HCS)},
  pages={1--43},
  year={2020},
  organization={IEEE Computer Society}
}

@article{dave2021hardware,
  title={Hardware acceleration of sparse and irregular tensor computations of ml models: A survey and insights},
  author={Dave, Shail and Baghdadi, Riyadh and Nowatzki, Tony and Avancha, Sasikanth and Shrivastava, Aviral and Li, Baoxin},
  journal={Proceedings of the IEEE},
  volume={109},
  number={10},
  pages={1706--1752},
  year={2021},
  publisher={IEEE}
}

@article{kim2015ramulator,
  title={Ramulator: A fast and extensible DRAM simulator},
  author={Kim, Yoongu and Yang, Weikun and Mutlu, Onur},
  journal={IEEE Computer architecture letters},
  volume={15},
  number={1},
  pages={45--49},
  year={2015},
  publisher={IEEE}
}

@article{liu2020neural,
  title={Neural sparse voxel fields},
  author={Liu, Lingjie and Gu, Jiatao and Zaw Lin, Kyaw and Chua, Tat-Seng and Theobalt, Christian},
  journal={Advances in Neural Information Processing Systems},
  volume={33},
  pages={15651--15663},
  year={2020}
}

@article{paszke2019pytorch,
  title={Pytorch: An imperative style, high-performance deep learning library},
  author={Paszke, Adam and Gross, Sam and Massa, Francisco and Lerer, Adam and Bradbury, James and Chanan, Gregory and Killeen, Trevor and Lin, Zeming and Gimelshein, Natalia and Antiga, Luca and others},
  journal={Advances in neural information processing systems},
  volume={32},
  year={2019}
}

@inproceedings{chen2022tensorf,
  title={Tensorf: Tensorial radiance fields},
  author={Chen, Anpei and Xu, Zexiang and Geiger, Andreas and Yu, Jingyi and Su, Hao},
  booktitle={European conference on computer vision},
  pages={333--350},
  year={2022},
  organization={Springer}
}

\end{document}